\documentclass[useAMS,usenatbib]{mn2e}
\usepackage[percent]{overpic}
\usepackage{times}
\usepackage{graphicx}
\usepackage{amsmath}
\usepackage{mathabx}
\usepackage{subfigure}
\usepackage{natbib}
\usepackage{txfonts}
\usepackage{journals}

\newcommand{\HII}{$\rm H~{\scriptstyle II}$}

\newcommand{\bsnr}{$\rm SNR(4650\,\AA)$}

\title[LSS-GAC: selection function]
{The selection function of the LAMOST Spectroscopic Survey of the Galactic Anticentre} 

\author[B.Q. Chen et al.]
{B.-Q. Chen,$^{1,2}$\thanks{E-mail:
bchen@ynu.edu.cn (BQC); x.liu@pku.edu.cn (XWL).}
 X.-W. Liu,$^{1,2,3}$\footnotemark[1]
 H.-B. Yuan,$^4$
 M.-S. Xiang,$^5$\thanks{LAMOST Fellow.}
 Y. Huang,$^{1,2}$\footnotemark[2] 
 C. Wang,$^2$
      \newauthor
H.-W. Zhang,$^2$
and Z.-J. Tian$^2$\footnotemark[2]
\\
$^{1}$South-Western Institute for Astronomy Research, Yunnan University, Kunming, Yunnan 650091, P.\,R.\,China\\
$^{2}$Department of Astronomy, Peking University, Beijing 100871, P.\,R.\,China\\
$^{3}$Kavli Institute for Astronomy and Astrophysics,
Peking University, Beijing 100871, P.\,R.\,China\\
$^{4}$Department of Astronomy, Beijing Normal University, Beijing 100875, P.\,R.\,China\\
$^{5}$National Astronomy Observatories, Chinese Academy of Sciences, Beijing 100012, P.\,R.\,China
 }

\begin{document}

\date{Accepted ???. Received ???; in original form ???}

\pagerange{\pageref{firstpage}--\pageref{lastpage}} \pubyear{2016}
\maketitle
\label{firstpage}

\begin{abstract}
We present a detailed analysis of the selection function of the LAMOST 
Spectroscopic Survey of the Galactic Anti-centre (LSS-GAC). LSS-GAC was 
designed to obtain low resolution optical spectra for a sample of more than 3 million 
stars in the Galactic anti-centre. The second release of value-added 
catalogues of the LSS-GAC (LSS-GAC DR2) contains stellar parameters, including radial velocity, 
atmospheric parameters, elemental abundances and absolute magnitudes deduced from
1.8 million spectra of 1.4 million unique stars targeted
by the LSS-GAC between 2011 and 2014.
For many studies using this database, such as 
those investigating the chemodynamical structure of the Milky Way, 
a detailed understanding of the selection function of the survey is indispensable.
In this paper, we describe how the selection function of the LSS-GAC can 
be evaluated to sufficient detail
and provide selection function corrections for all spectroscopic measurements 
with reliable parameters released in LSS-GAC DR2. 
The results, to be released as new entries in the LSS-GAC
value-added catalogues, can be used to correct the selection effects of 
the catalogue for scientific studies of various purposes. 
\end{abstract}


\section{Introduction}
Large-scale spectroscopic surveys of Galactic stars, e.g. the
Sloan Extension for Galactic Understanding and Exploration 
(SEGUE; \citealt{Yanny2009}) 
and the LAMOST Experiment for Galactic Understanding and 
Exploration (LEGUE; \citealt{Deng2012, Zhao2012}),
are opening a new window for the study of the formation and evolution of 
the Milky Way galaxy in great detail. However,
unlike photometric surveys that yield, in general,  complete samples of objects to a given 
limiting magnitude, time-consuming spectroscopic surveys 
often have to select targets, and are unavoidably affected 
by the various potential target selection effects. Bias arises from 
the target selection, the observation,
data reduction and processes determining parameters. 
To understand the relationship between 
a spectroscopic sample of stars with a reliable estimation of parameters 
and the parent stellar population, 
one needs to study and account for the selection function.

Many authors have made efforts to characterise the selection function 
of spectroscopic samples from various completed or on-going surveys. 
\citet{Cheng2012} determine the selection function of 
a sample of SEGUE main-sequence turn-off stars. 
\citet{Schlesinger2012} study and correct for the 
various selection biases of SEGUE G and K dwarfs.  
Selection effects are also considered in 
\citet{Bovy2012} and \citet{Liu2012} for SEGUE G dwarfs for different purposes. 
\citet{Nidever2014} characterise the selection
effects of the Apache Point Observatory Galactic Evolution 
Experiment (APOGEE; \citealt{Majewski2015}) red clump stars. 
More recently, \citet{Stonkute2016} discuss
the selection function of Milky Way field stars targeted by the Gaia-ESO survey \citep{Gilmore2012},
and \citet{Wojno2016} describe in detail the selection function of the 
Radial Velocity Experiment (RAVE; \citealt{Steinmetz2006}) survey.

In this paper, we take efforts to 
analyze the selection function of the LAMOST 
spectroscopic Survey of the Galactic Anticentre 
(LSS-GAC; \citealt{Liu2014, Liu2015,Yuan2015}).
LSS-GAC is a major component of LEGUE. 
It was initiated in October, 2012, following a year-long Pilot Survey.
It aims to observe $\sim$ 3 million stars of all colours
and magnitudes of $r$ $\lesssim$17.8\,mag (18.5 for a limited number of fields)
in a large (3,400\,deg$^2$) and continuous sky area 
centred on the Galactic anti-centre (GAC). 
The survey should allow us to obtain a deeper 
understanding of the structure, formation and evolution of the Milky Way disk(s), 
and of the Galaxy as a whole.

Data yielded by the  LSS-GAC survey are available from the LAMOST official data
releases, such as LAMOST DR1 \citep{Luo2015}. 
The official data releases include stellar
spectra and stellar parameters derived with the LAMOST Stellar
parameter Pipeline (LASP; \citealt{Wu2011, Wu2014}).
In addition, there are public releases of LSS-GAC 
value-added catalogues, the LSS-GAC DR1
\citep{Yuan2015} and LSS-GAC 
DR2\footnote{http://lamost973.pku.edu.cn/site/data} \citep{Xiang2016}. 
The LSS-GAC DR1 contains radial velocities and stellar atmospheric parameters 
derived with a different stellar parameter pipeline, the LAMOST Stellar Parameter Pipeline at Peking University 
(LSP3; \citealt{Xiang2015a}), for LAMOST spectroscopic observations 
between September, 2011 and June, 2013. The catalogue also presents additional
information, including multiband photometry  
and proper motions collected from various databases,  
as well as extinction, distance and orbital parameters 
deduced with a variety of techniques. For LSS-GAC DR2, in addition to the above information, 
$\alpha$-element abundances, C and N abundances,
and absolute magnitudes derived from the improved LSP3 \citep{Xiang2016b}
are also provided for LAMOST spectroscopic observations 
between September, 2011 and June, 2014.

In this paper we present a detailed study of the selection function
of LSS-GAC based on its most recent data release, 
LSS-GAC DR2 \citep{Xiang2016}, 
to facilitate broad and robust usage of this
publicly-available database.
There have already been several studies trying to 
characterise the selection function of stars targeted by LSS-GAC.
\citet{Rebassa2015} have discussed the selection function of a small sample of 
white dwarfs identified in an early stage of LSS-GAC in a study aimed to  
determine the mass function of Galactic white dwarfs. 
\citet{Xiang2015} have carried out a detailed analysis
of the selection effects of 
LSS-GAC F-type turn off stars that they used to determine
metallicity gradients of the Milky Way disk. \citet{Liu2017} take 
the selection effects of LAMOST K giants 
into account when deriving the stellar number density distribution.

However, a comprehensive analysis of the selection function of LSS-GAC is still lacking. 
The earlier efforts of \citet{Rebassa2015}, \citet{Xiang2015}, and \citet{Liu2017} 
all concentrate on specific samples selected from LSS-GAC.
In addition, \citet{Liu2017} determine the selection 
function of stars on the basis of the individual LAMOST plates, which 
have a field of view (FoV) of $\sim$20\,deg$^2$. The approach is not suitable 
for LSS-GAC, which selects targets based on  boxes 
of $\sim$ 1\,deg$^2$ in sky area. Given the steep stellar 
number density gradients with latitude near the Galactic plane, 
the selection function in different parts of a given LAMOST 
plate near the plane would be quite different.  
Furthermore, LAMOST is equipped with 16 spectrographs
that have different throughputs, decreasing in general 
with distance from the field centre \citep{Yuan2015}. 
Clearly, such variations of the selection function can not be ignored.
\citet{Xiang2015} improve the work by determining 
the selection function spectrograph by spectrograph. 
However in evaluating the selection function, 
they have combined stars in a given spectrograph
targeted by all LAMOST plates that share the same central star.
Different plates are usually observed under different weather conditions 
including transparency,  seeing and lunar phase, 
and are thus likely to have different limiting magnitudes.
The selection function is expected to differ significantly amongst 
different plates. Furthermore, in some rare cases, although two plates target the 
same field, the sky areas targeted by the individual spectrographs
of the two plates actually differ. Thus evaluating the selection function by
combining data from the different plates is inappropriate. 
 
In this work, we discuss in detail the selection function 
of spectroscopic measurements of stars catalogued
in the LSS-GAC DR2 and give a robust way to evaluate the selection bias, by 
considering as many effects as possible. Mock data are used to test our
technique for the selection function evaluation.
The paper is organised as follows. In \S{2} we introduce briefly
the LSS-GAC, including the target selection algorithm and the LSS-GAC DR2. 
In \S{3} we describe how we evaluate the selection function of LSS-GAC. 
In \S{4} we test our algorithm using mock data. 
We discuss the applications of our results in \S{5} and 
summarise in \S{6}.

\section{LSS-GAC}

\begin{figure*}
    \centering
\includegraphics[width=0.88\textwidth]{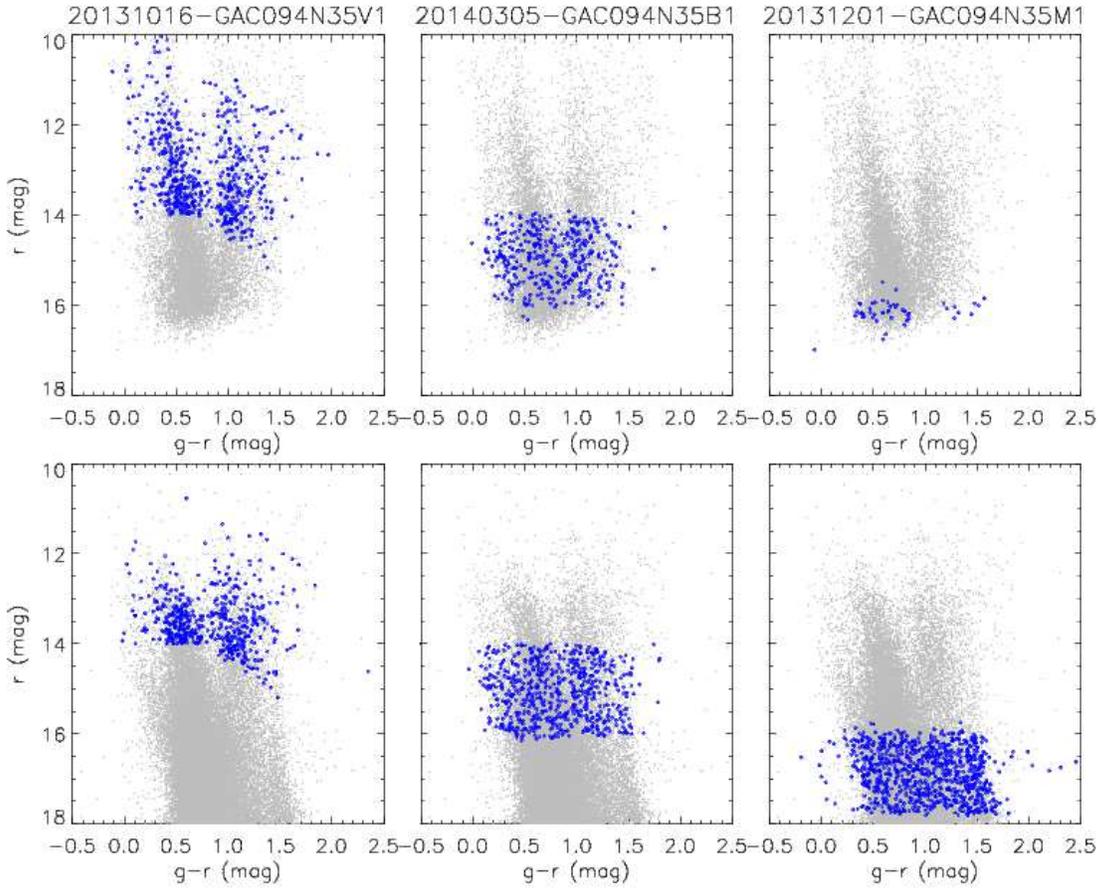}
  \caption{Target selection for three LSS-GAC example plates
all centred on RA~=~94.39059\degr and Dec~=~35.14920\degr.
The grey dots represent all stars in the clean photometric samples of the field.
The blue dots represent all those selected and observed objects by the three LSS-GAC plates.
 The upper and bottom rows are based on the APASS DR9 and 
XSTPS-GAC photometric catalogues, respectively. 
The left, middle and right columns show target selection for the
VB, B and M plates, respectively. The observational dates and IDs of the individual plates are labeled at 
the top of the three columns.}
  \label{scmd}
\end{figure*}

LSS-GAC contains three different components, the main, the M31/M33 and 
the VB surveys. The main survey aims to observe about 3 million stars in a contiguous sky area 
towards the GAC (150\degr $< l< $ 210\degr, and $-$30\degr $< b< $30\degr). 
The M31/M33 survey observes all kinds of interesting targets 
in the vicinity fields of M31 and M33 within the reach of  LAMOST,
including supergiants, massive star clusters, planetary nebulae, \HII\ regions, 
as well as background QSOs, galaxies and foreground Galactic stars.
The VB survey is designed to observe
very bright (VB) stars  (9 $< r <$ 14\,mag) in sky areas
accessible to LAMOST ($-$10$\degr < {\rm Dec} <$ 60\degr) for time of non-ideal 
observing conditions such as in bright/grey lunar nights. 
  
In this section we will give a brief introduction to the LSS-GAC survey and its 
most recent data release, LSS-GAC DR2.
\citet{Liu2014} introduce the survey design
and scientific motivations of LSS-GAC. \citet{Yuan2015} present the 
target selection and the LSS-GAC DR1. \citet{Liu2015} 
give a review of the early scientific results.  
\citet{Xiang2015a, Xiang2015b} and \citet{Xiang2016b} describe the data reduction 
of LSS-GAC and \citet{Xiang2016} present the LSS-GAC DR2.  

\subsection{Target selection}

Four different types of survey plates, namely, very bright (VB) , 
bright (B), median bright (M) and faint (F) plates, are designed for LSS-GAC. 
They are defined by different $r$-band magnitude ranges. 
Usually, VB plates target stars of $r < 14$\,mag. 
B, M and F plates target stars of $14 < r \le m_1$\,mag, $m_1 < r \le m_2$\,mag and 
$m_2 \le r < 18.5$\,mag, respectively. Here $m_1$ and $m_2$ are the border magnitudes
separating B, M and F plates and differ slightly for different regions in the sky \citep{Yuan2015}.
Typical values of $m_1$ and $m_2$ are 16.3 and 17.8\,mag, respectively.
Except for some VB plates, LSS-GAC targets are selected from the photometric catalogues of
Xuyi Schmidt Telescope Photometric Survey of the Galactic Anticentre 
(XSTPS-GAC; \citealt{Liu2014, Yuan2015}). 

XSTPS-GAC surveys an area of approximately 7,000\,deg$^2$ 
in the GAC area, including the M31/M33 region, using
the Xuyi 1.04/1.20m Schmidt Telescope.
It collects images in SDSS $g$, $r$ and $i$ bands.
XSTPS-GAC catalogues about
one hundred million stars down to a limiting magnitude of
$r$ $\sim$ 19.0\,mag (10$\sigma$).  The photometric systematic uncertainties
of XSTPS-GAC are estimated to be smaller than 0.02\,mag (Yuan
et al., in preparation), and the
uncertainties of resulted RA and Dec are about 0.1\,arcsec \citep{Zhang2014}. 

The basic strategy of the LSS-GAC target selection is to uniformly and randomly select stars
from the colour-magnitude diagrams \citep{Yuan2015}. A brief summary of 
the target selection procedure for the LSS-GAC B, M and F plates is as follows:
\begin{enumerate}
 \item The XSTPS-GAC photometric catalogue is used to generate a clean
sample of targets for LSS-GAC, excluding stars that are either poorly detected, 
badly positioned, flagged as galaxies or star pairs, 
or contaminated by bright neighbours or by sky background.
\item The whole survey area is divided into boxes of 1\,deg side in
RA and Dec. For stars in each box, ($r$, $g - r$) and ($r,$ $r - i$)
Hess diagrams are constructed from the clean sample. 
Stars of extremely blue colours, $(g - r)$ or $(r - i)$ $<-$0.5\,mag, and
of extremely red colours, $(g - r)$ or $(r - i)$ $ >$ 2.5\,mag, are first
selected. 
\item Stars in the remaining colour space are  then
sorted in magnitude from bright to faint. 
$m_1$ and $m_2$ are set to the faint end magnitudes of the first 40\% and
80\% sources, respectively. 
\item Stars of B, M and F plates are selected and assigned priorities, in batches
of 200 stars per deg$^2$, with a Monte Carlo (random) approach. 
\item The field centres of the individual LAMOST plates are defined.
All LAMOST plates must be centred on bright stars ($\lesssim 8$\,mag) such that 
the LAMOST active optics can operate.
\item For each plate, the SSS software \citep{Luo2015} is used to allocate fibres to 
the selected stars.
\end{enumerate} 

The target selection of VB plates is slightly different from B, M and F plates. 
Within the XSTPS-GAC footprint,
all stars of $r \le$ 14.0\,mag from XSTPS-GAC
and all stars of 9 $\le J  \le$ 12.5\,mag from Two Micron 
All-Sky Survey (2MASS; \citealt{Skrutskie2006})
are selected as potential targets with equal priorities. Outside the XSTPS-GAC
footprint, all stars of 10.0 $\le b1 \le $ 15.0\,mag, or 10.0 $\le b2 \le$ 15.0\,mag, 
or 9.0  $\le r1 \le$  14.0\,mag, or 9.0  $\le r2 \le$  14.0\,mag, or
8.5 $\le i \le$ 13.5mag from PPMXL \citep{Roeser2010} 
and stars of 9 $\le J \le$ 12.5\,mag from 2MASS  
are selected as potential targets with equal priorities.

\subsection{Observation, data reduction and the LSS-GAC DR2}

The five-year long Phase~I LAMOST Regular Surveys were initiated in
October 2012, following the one-year long Pilot Surveys. 
In each year, a sufficient number of the LSS-GAC plates 
are planned in advance for the observation. 
The main and the M31/M33 survey plates are observed in dark/grey
nights. Typically 2 -- 3 exposures are obtained for each plate,
with typical integration time per exposure of 600 -- 1200\,s,
1200 -- 1800\,s, 1800 -- 2400\,s for B, M and F plates, respectively, 
depending on the weather. 
The seeing varies between 3 -- 4\,arcsec for most plates, 
with a typical value of about 3.5\,arcsec. 
The VB plates, typically observed
with 2~$\times$~600\,s, are observed in bright nights or nights of
poor observing conditions.

In total, 314 plates (194 B + 103M + 17 F) for the LSS-GAC
main survey, 59 plates (38 B + 17 M + 4 F) for the M31/M33
survey and 682 plates for the VB survey have been observed by June, 2014.
The raw spectra are first processed with the LAMOST 2D
pipeline \citep{Luo2012, Luo2015} to extract the 1D spectra.
The resultant 1D spectra are then processed with LSP3 to
obtain radial velocities, basic atmospheric parameters ($T_{\rm eff},~{\rm log}\,g$ and [Fe/H]) 
as well as [$\alpha$/Fe],  [C/H] and [N/H] abundance ratios, and 
absolute magnitudes $M_V$ and $M_{K_{\rm s}}$.
The resultant parameters serve as the core
data of the LSS-GAC DR2 \citep{Xiang2016b}.

The most recently published LSS-GAC DR2 contains information derived 
from 1.8 million spectra of 1.4 million unique stars
that have a spectral signal to noise ratio (SNR) at 4650\AA\
higher than 10, collected for the LSS-GAC main, 
M31/M33 and VB surveys since 2011 September until 2014 June. 
LSS-GAC DR2 provides additional
information of the individual targets, including the observing conditions, 
and absolute magnitudes, values of interstellar 
extinction, distances, and orbital parameters  derived from the basic 
parameters using a variety of techniques.

\section{The selection function}

In this paper, we consider the selection function to be the relation between 
a spectroscopic sample selected from the LSS-GAC 
value-added catalogue with robust determinations for (certain) stellar parameters and
the underlying (statistically complete to a given limiting magnitude) photometric sample. 
Generally, the selection effects of a selected LSS-GAC spectroscopic sample 
are due to the following two parts:  (1) the LSS-GAC target selection algorithm,
and  (2) the observation, data reduction and parameter determination processes. 
We define the 
selection function, $S$, as the probability of a star which is selected, observed and 
ends up as a valid entry in the LSS-GAC value-added catalogue. 
The selection function $S$ can then be divided into two
parts, namely, $S_1$, the probability of a star in a given colour-magnitude bin
that is selected and gets observed by LAMOST, and
$S_2$, the probability of a LAMOST spectrum of a star in a given colour-magnitude bin
that is capable of delivering robust stellar parameters. 

In the current work, we discuss the selection function $S$ of LSS-GAC mainly based on 
the photometric data of XSTPS-GAC, as most of the LSS-GAC targets
are selected from XSTPS-GAC. Limited by the sky coverage and bright star saturation 
of XSTPS-GAC, some VB stars are selected  from PPMXL and 2MASS.
For those plates we adopt the AAVSO Photometric All-Sky Survey (APASS; \citealt{Henden2016})
DR9 catalogue to determine the selection function of the spectroscopic measurements.
The APASS survey is conducted in five filters, including the 
Johnson $B$ and $V$, and Sloan $g, ~r,~ i$ bands. 
It covers the entire sky and is valid for magnitude range $7 < r < 17$\,mag. It thus 
serves as an excellent photometric catalogue
for the LSS-GAC VB targets. The APASS DR9 contains photometric data of $\sim$ 61 million 
measurements covering about 99\,per\,cent of the sky. We remove all the repeated measurements
in APASS DR9 (about 7\,per\,cent) and keep only those with the smallest
photometric uncertainties for the individual stars.
We require that all stars should have detections in $g$ and $r$ bands in 
both the XSTPS-GAC and APASS DR9 catalogues.
In Fig.~\ref{scmd}, we show the colour-magnitude diagrams (CMD) 
for stars targeted by three LSS-GAC plates, VB, B and M each, with all stars from
the XSTPS-GAC and APASS DR9 catalogues overplotted. 
The Figure shows clearly how LSS-GAC targets are selected 
from the photometric catalogues.

For all the spectroscopic measurements with reliable parameters released in 
LSS-GAC DR2, values of selection function 
are calculated based on the XSTPS-GAC and on the APASS
photometric catalogue separately. For stars falling inside the XSTPS-GAC footprint 
and having magnitudes in the range of 13.0 $<$ $r$ $<$ 18.0\,mag, the
selection function values  calculated with the XSTPS-GAC catalogue are adopted. For 
stars falling outside the XSTPS-GAC footprint and having magnitudes 
in the range of 7.0 $<$ $r$ $<$ 15.5\,mag, or stars falling inside the  
XSTPS-GAC footprint but having magnitudes 
in the range of 7.0 $<$ $r$ $<$ 13.0\,mag, the
selection function values calculated by the APASS DR9 catalogue are adopted.

\subsection{Selection effect due to target selection}

\begin{figure*}
    \centering
\includegraphics[width=0.78\textwidth]{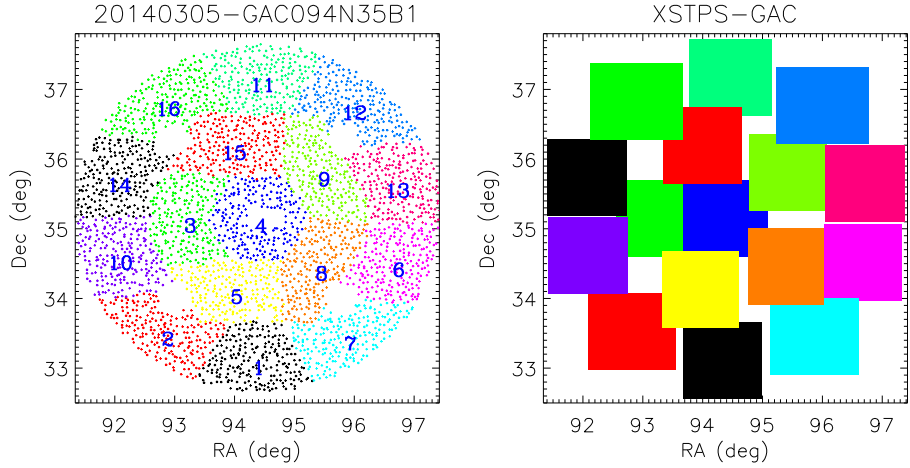}
  \caption{Spatial distribution of stars targeted by an example LSS-GAC plate 20140305-GAC094N35B1
(left) and the layout of the 16 rectangular boxes that are used to define the parent photometric samples from the
XSTPS-GAC catalogue for the 16 spectrographs of LAMOST (right; see text for detail). 
In the left panel, stars targeted by different spectrographs are
plotted with different colours. The spectrograph IDs are labelled. In the right panel,
the rectangles are plotted with the same colour  
as the stars in the spectrographs they represent on the left.}
  \label{sd}
\end{figure*}

\begin{figure*}
    \centering
\includegraphics[width=0.88\textwidth]{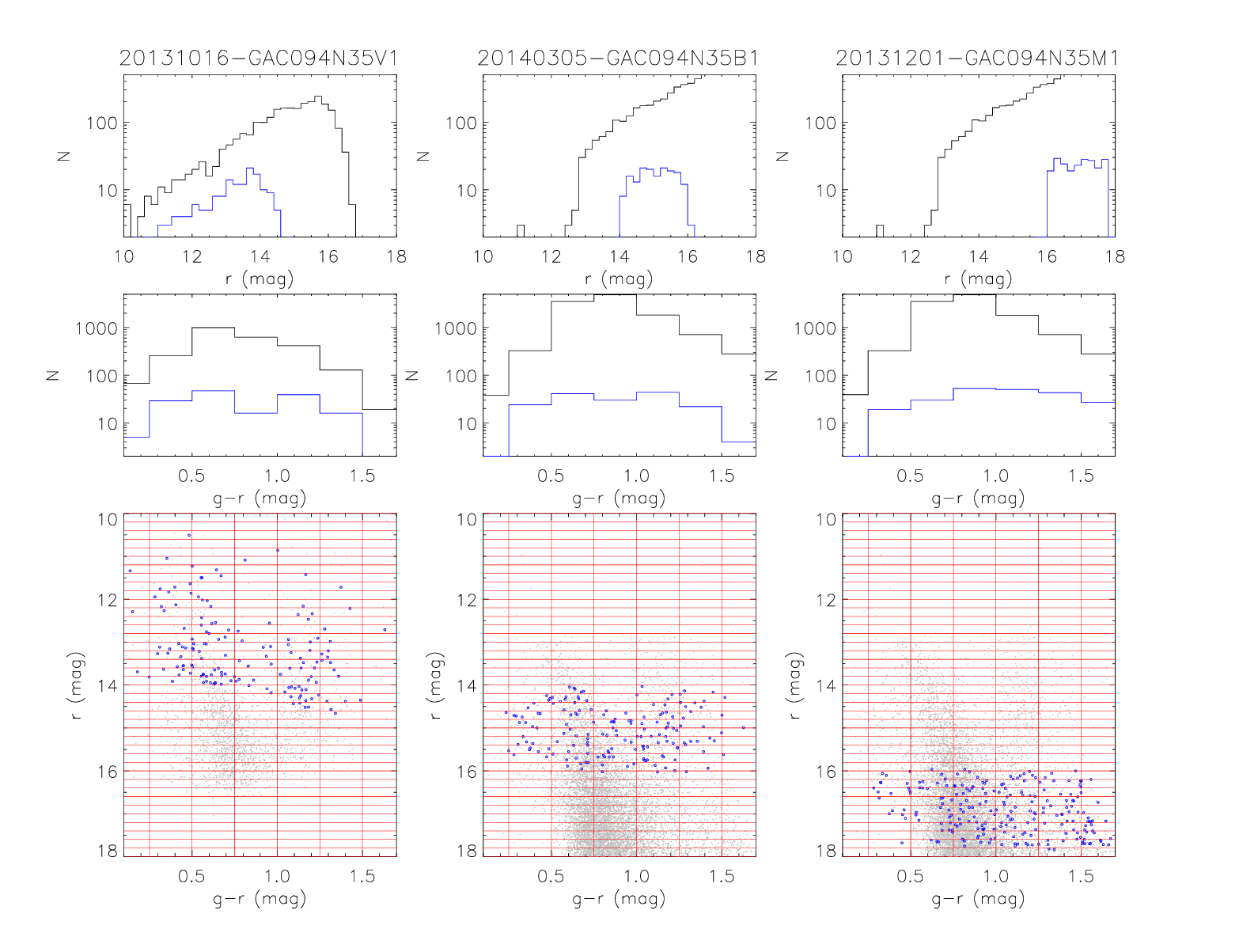}
\caption{Distributions in CMDs of stars targeted by Spectrograph~\#1 
of three example plates (blue dots and histograms) and of all stars in 
the representative rectangular box, for the case of using the APASS (left panel, a VB plate) or 
the XSTPS-GAC (middle panel, a B plate and right panel, an M plate) photometric catalogues.
Red lines in the bottom panels show the grid of bins for $S_1$ evaluation.
The observational dates and IDs of the three example plates are 
labeled at the top of the three columns.}
  \label{sacmd}
\end{figure*}

\begin{figure*}
    \centering
\includegraphics[width=0.68\textwidth]{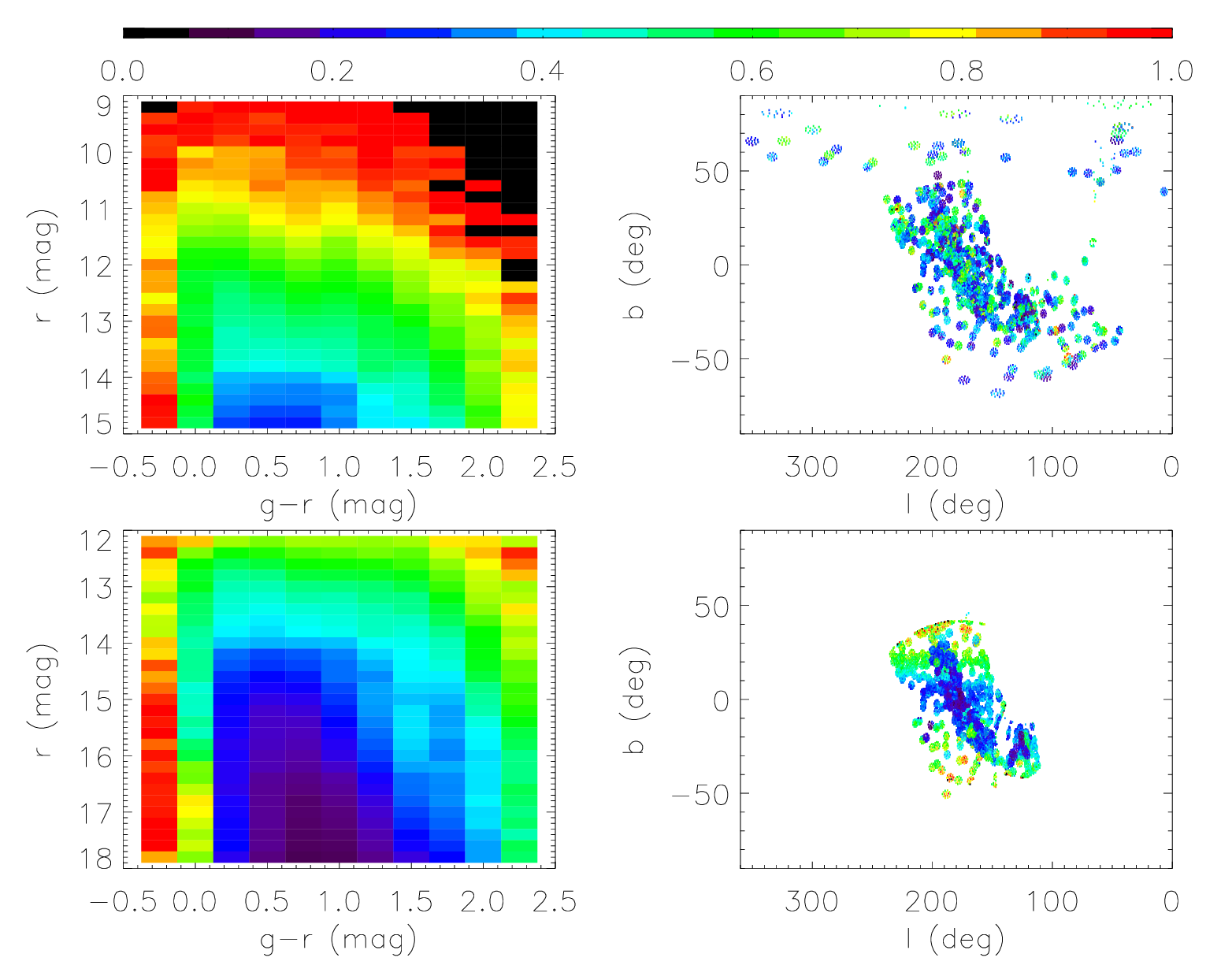}
\caption{Distributions of mean values of $S_1$ in CMDs (left panels) 
and in Galactic longitude-latitude ($l~,b$) plane (right panels)
for stars targeted by all spectrographs of all plates included in the LSS-GAC DR2. 
The upper and bottom panels represent 
respectively results obtained using the APASS and XSTPS-GAC photometric catalogues.}
  \label{f1d}
\end{figure*}

We first collect all spectroscopic measurements in LSS-GAC. For those measurements, 
the selection effects come from the LSS-GAC target selection algorithm, i.e. $S_1$. 
We calculate $S_1$ for the individual spectrographs of each LSS-GAC plate.  
In Fig.~\ref{sd}, we show the spatial distribution in the sky for
the measurements in different spectrographs of a given LSS-GAC plate. 
The boundaries of the individual spectrographs 
are irregular and the areas covered by them differ from each other.  
To make a robust comparison between
the LSS-GAC targeted sample and the parent photometric sample, 
we consider each plate to be composed of rectangles, 
roughly centred on each spectrograph. The area of the rectangle 
is chosen to cover the same area in the sky as the corresponding spectrograph.
Thus the stellar distribution in the rectangular area 
would be the same as that in the corresponding spectrograph, 
which can be used to calculate $S_1$. 
The centre of the rectangular area is set to the mean position of all stars  
targeted in the spectrograph. The size in Declination 
of the rectangular area is set to 1.1\degr, while the
size in Right Ascension is set  to $\Omega/(1.1\cos\delta)$, where $\Omega$ is
the size of the spectrograph (unit in deg$^2$ ) and 
$\delta$ the central Declination of the spectrograph.
We show in the right panel of Fig.~\ref{sd} the rectangular areas for 
the individual spectrographs.

A LAMOST plate has a circular FoV of  $\sim$20\,deg, 
covering a diameter of 5\,deg. Thus the individual spectrographs have an average
area of $\sim$ 1.2\,deg$^2$, comparable to the size of the box that we used for 
target selection (1\degr\ $\times$ 1\degr). 
According to the LSS-GAC target selection strategy, 
for the individual spectrographs of LAMOST, the 
probability of a star in a given colour-magnitude bin that is selected and gets observed by LAMOST, $S_1$,
can be calculated as,
\begin{equation}
  S_1= \dfrac{N_{\rm LAMOST}({\rm sp},C,M)} {N_{\rm phot.}({\rm sp},C,M)},
\end{equation}
where ${N_{\rm LAMOST}({\rm sp},C,M)}$ and ${N_{\rm phot.}({\rm sp},C,M)}$ are
respectively the number of all LAMOST measurements and the number of all photometric stars 
in a given colour $C$ and magnitude $M$ bin for Spectrograph sp
of a LSS-GAC plate. For both the calculation with XSTPS-GAC and APASS
photometric catalogues, $C$ and $M$ corresponds to $g-r$ and $r$, respectively. 
We adopt a colour and magnitude bin-size of $\delta (g-r)$~=~0.25\,mag and $\delta r$~=~0.2\,mag. 
Fig.~\ref{sacmd} shows the colour and magnitude distributions 
of all targets observed with LAMOST and of
the underlying photometric samples, along with the grid we use to calculate $S_1$, 
for Spectrograph~\#1 of three example  plates. 

The distributions of averaged values of $S_1$ in the
CMD and in the Galactic longitude-latitude $(l, b)$ plane for targets observed with all
spectrographs of all plates included in the LSS-GAC DR2 are shown in Fig.~\ref{f1d}
for the cases of selection with APASS and selection with XSTPS-GAC.
In general, brighter stars have higher values of $S_1$ than the fainter ones, and 
stars of extreme colours have higher values of  $S_1$ 
than those of medium colours. The result is consistent with 
the strategy of LSS-GAC target selection, i.e., selecting stars uniformly and randomly 
from the CMDs. Stars of fainter magnitudes or of medium colours are much more numerous than 
those of brighter magnitudes or of extreme colours, so they  
have lower probabilities to get observed, i.e. smaller values of $S_1$.
Spectrographs (of plates) of high Galactic latitudes have higher average values of 
$S_1$ than those of lower latitudes. Again, this is simply due to the steep 
decline of stellar number density (to a given limited magnitude) with latitude.

Note that there exists substantial overlaps
between adjacent LSS-GAC plates. In addition, often there are more
than two plates targeting the same field, covering exactly the same sky area.
Nevertheless, it is extremely important to bear in mind that values of 
$S_1$ calculated here are for stars targeted by individual (plate) observations 
of LAMOST. For any follow-up scientific applications, e.g. to derive the underlying 
stellar number density of a given age from the LAMOST spectroscopic sample with 
robust stellar parameter determinations, results from the individual observations can 
only be combined after correcting for the selection effects.

\subsection{Selection effect due to  observation, data reduction and parameter determination}

\begin{table*}
   \centering
  \caption{Probabilities $S_1$, $S_2$ and $S$ for measurements catalogued  in the LSS-GAC DR2.}
  \begin{tabular}{lrrrrrrrrr}
  \hline
  \hline
Spec\_id & Date & Plate & Spectrograph & RA &  Dec &  $ S_1$
&  $S_2$ &  $S$ & Notes$^{a}$ \\
  \hline
 20110921-PM1-01-003 & 20110921   & PM1   &  01 & 12.90091  & 35.61838 &   0.100 &   0.600 &   0.060  & 1 \\
 20110921-PM1-01-005 & 20110921   & PM1   &  01 &13.23111  & 35.65357 &   0.154 &   0.833 &   0.128  & 1 \\
 20110921-PM1-01-006 & 20110921   & PM1   &  01 &12.98870  & 35.79328 &   0.182 &   1.000 &   0.182  & 1 \\
 20110921-PM1-01-007 & 20110921   & PM1   &  01 &13.24852  & 35.89709 &   0.750 &   1.000 &   0.667  & 1 \\
  ... & ...   & ...  &  ...  & ... &   ... &   ...  &   ...  & ... & ... \\
 20110921-PM1-01-011 & 20110921   & PM1   & 01  &13.18381  & 35.87553 &   0.062 &   1.000 &   0.062  & 2 \\
 20110921-PM1-01-013 & 20110921   & PM1   &  01 &12.93216  & 35.62636 &   0.571 &   0.750 &   0.429  & 2 \\ 
  ... & ...   & ...  &  ...  & ... &   ... &   ...  &   ...  & ...  & ...\\
  \hline
\end{tabular}\\
\begin{flushleft}
{$^a$ 1: The probabilities calculated with the XSTPS-GAC photometric catalogue. 2: The probabilities calculated with the APASS photometric catalogue.}
\end{flushleft}
\end{table*}

\begin{figure*}
    \centering
\includegraphics[width=0.98\textwidth]{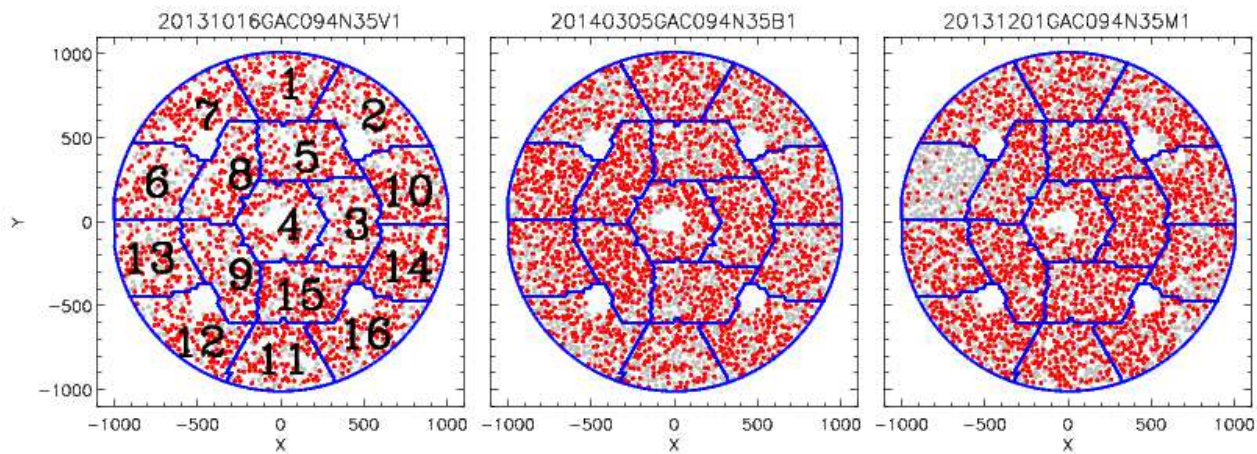}
  \caption{Spatial distributions of stars in the LAMOST focal plane (X, Y). 
The stars with reliable parameters determined with LSP3 are denoted with red dots. 
All stars selected and observed by LAMOST are denoted with grey dots. 
The three panels are corresponding to three example plates
 (a  VB, a B and an M plate  from left to right).
The boundaries of the individual spectrographs are delineated by blue lines
with IDs marked. The observational dates and IDs of the three plates are labelled on the 
top of the panels.  }
  \label{lstp}
\end{figure*}

\begin{figure*}
    \centering
\includegraphics[width=0.88\textwidth]{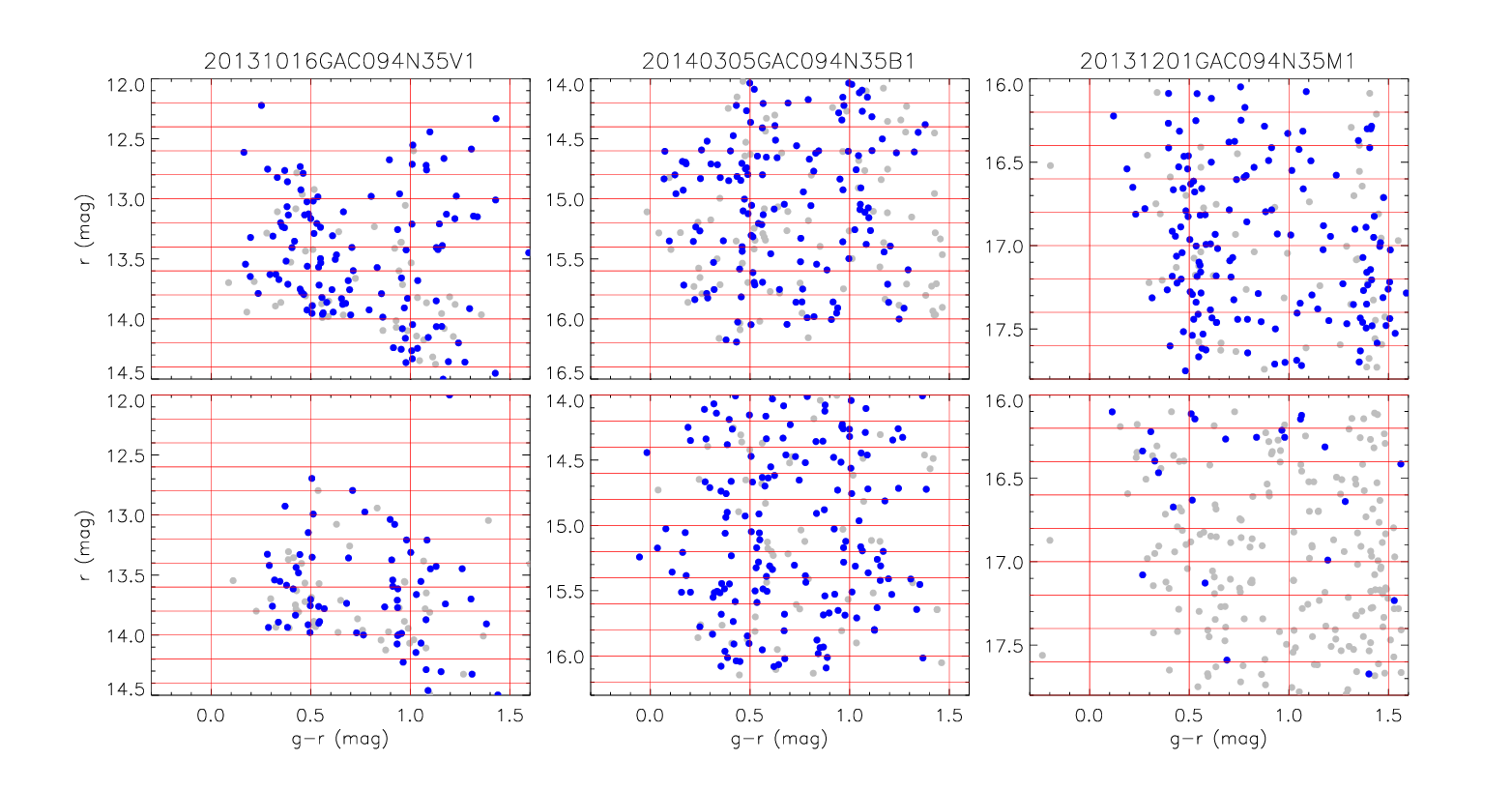}
  \caption{CMDs for stars observed with two selected spectrographs
of three example plates. In each panel all stars selected and get observed with 
LAMOST (grey dots) and those
with reliable parameters determined with LSP3 (blue dots) are shown. Values of colour $g-r$ 
and magnitude $r$ are from the XSTPS-GAC catalogue.  
Red lines delineate the grid of bins for $S_2$ evaluation.
The observational dates and IDs of the plates are labelled on the 
top of the columns.  The upper and bottom panels are for 
Spectrograph~\#15 and \#6 of those plates, respectively.  }
  \label{cmdsp}
\end{figure*}

\begin{figure*}
    \centering
\includegraphics[width=0.68\textwidth]{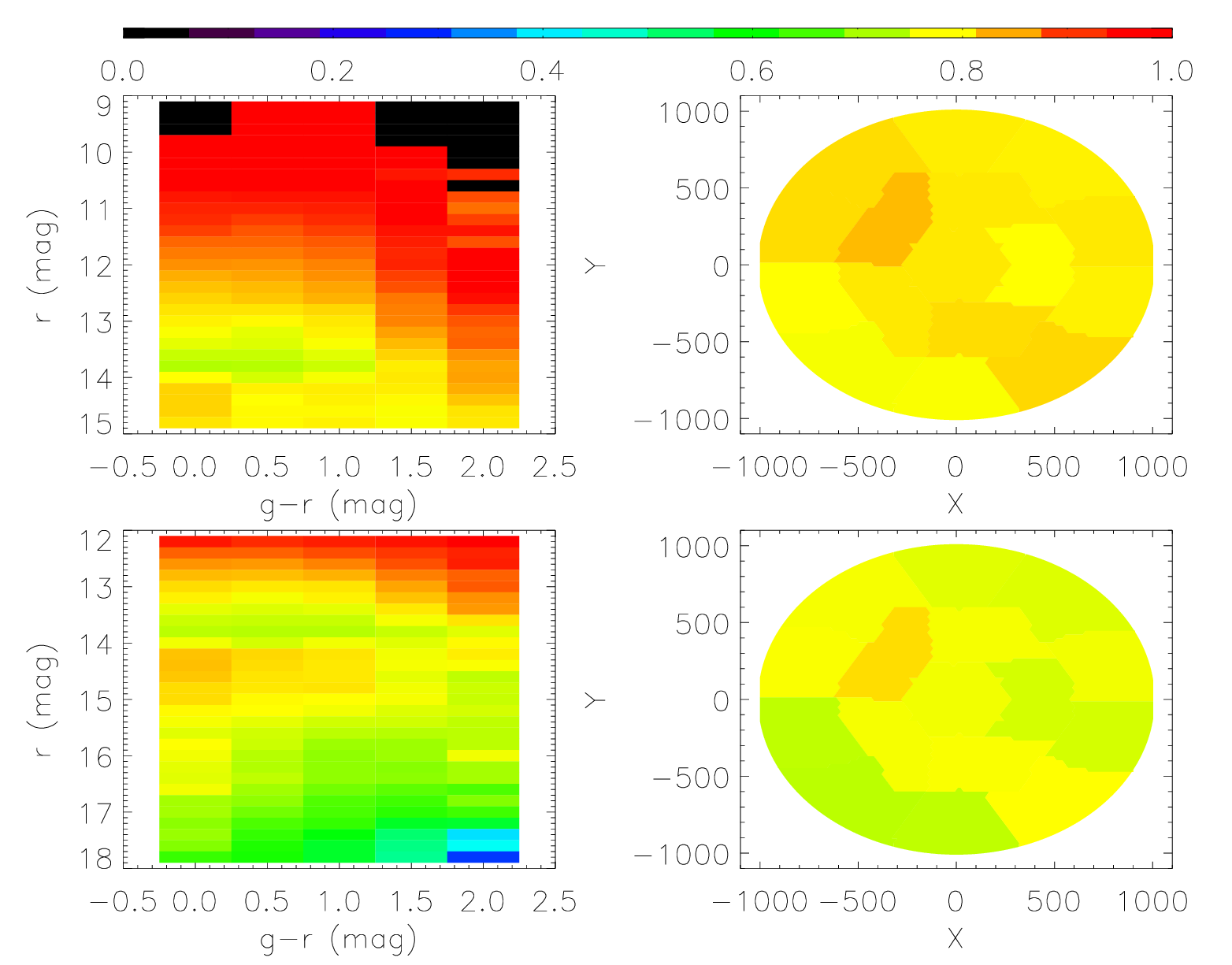}
  \caption{Distribution of averaged values of 
 $S_2$ in the colour and magnitude $(C,~M)$ space (left panels) 
and in LAMOST focal plane ($X~,Y$) (right panels) for
all spectroscopic measurements with robust stellar parameters in LSS-GAC DR2. 
The top and bottom panels represent 
the results obtained using the APASS and XSTPS-GAC catalogues, respectively.}
  \label{f2d}
\end{figure*}

$S_1$ represents the bias induced by the LSS-GAC target selection algorithm. Accounting for $S_1$ 
will eliminate the discrepancy between the number of stars targeted by LAMOST
and the number of stars in the parent photometric catalogue. 
However, there are additional effects that determine whether a spectroscopic measurement of 
star targeted by LAMOST ends up in the 
resultant spectroscopic catalogue with robust parameters.
This is related to the quality of the observation, data reduction, and parameter determination.
Robust estimates of stellar parameters, including radial velocities and basic atmospheric parameters, 
plus any additional information, such as values of interstellar extinction and 
distances  can only be deduced from spectra of sufficient quality.
Even for spectra of good quality, the currently developed stellar parameter pipelines 
(e.g. LSP3) may fail to deliver usable parameters because of lack of suitable analysis tools. 
This is the case for stars of extremely hot or cool colours. Thus one needs another quantity to 
account for this selection effect. 

The precision of stellar parameters deduced from the spectra are mainly 
determined by the quality of spectra. The requirements are however 
different for different parameters. For example,
robust radial velocities can be deduced for LAMOST spectra of \bsnr$>$5; while 
[$\alpha$/Fe] ratios can only be used for spectra of \bsnr$>$20 \citep{Xiang2016}.
Thus the selection criteria to build a spectroscopic sample with 
 `robust' parameters are different for different applications.
In the current work, we consider a commonly used sample 
selected from the  LSS-GAC DR2, following the 
recommendation of \citet{Xiang2016}. We select stars that have,
\begin{enumerate}
   \item snr\_b $>$ 10 for LAMOST spectra of good quality;
  \item moondis $>$ 30 to avoid moonlight contamination;
  \item vr\_flag $\le$ 6 and Teff $>$ 0 for robust radial velocity and atmospheric parameters;
  \item satflag $=$ 0 to avoid CCD saturation;
  \item brightflag $=$ 0 to eliminate contaminations by nearby bright stars;  
  \item deadfiber $=$ 0 to reject the spectra from LAMOST bad fibres;
\end{enumerate}  
The above criteria are the basic requirements for robust radial velocity and basic 
atmospheric parameters. We compare in Fig.~\ref{lstp}  the distributions of stars 
observed with LAMOST and those with reliable parameters in the LAMOST focus physical plane ($X,~Y$)  
for three example plates. A similar comparison for two example spectrographs 
of the above three example plates is given in Fig.~\ref{cmdsp} 
in the colour-magnitude ($C,~M$) plane. It is clear that 
the probabilities of stars with reliable parameters depend on the spectrographs, 
plates with which they are observed, as well as on
colours and magnitudes of the stars themselves.

As described above, whether a LAMOST spectrum 
is capable of delivering reliable stellar parameters
could depend on many factors. 
The most important factor is of course the SNR of the spectrum that depends 
on the brightness of the target, the exposure time and the observation conditions. The individual 
spectrographs of LAMOST also  have different throughputs \citep{Yuan2015}. 
Finally, the current implemented version of LSP3 pipeline uses the 
blue-arm spectra for parameter estimation. Thus stars of blue colours are more 
likely to yield spectra of sufficient SNRs and thus 
reliable stellar parameters than stars of red colours, either intrinsically red or 
heavily reddened by the interstellar dust grains. 
The quantity to account for all the above effects, $S_2$,
defined above as the probability of a LAMOST spectrum of a star in a given 
colour-magnitude bin that is capable of delivering robust stellar parameters, can be calculated as,
\begin{equation}
  S_2=\dfrac{N_{\rm PARAM}({\rm sp},C, M)}{N_{\rm LAMOST}({\rm sp},C, M)},
\end{equation} 
where ${N_{\rm PARAM}({\rm sp},C, M)}$ and ${N_{\rm LAMOST}({\rm sp},C, M)}$
are respectively the number of stars having reliable parameters in a colour-magnitude bin and 
the number of all stars in the same colour-magnitude bin that are selected and 
observed  by LAMOST in Spectrograph sp of a given  plate. Again,
for both selection with the XSTPS-GAC and APASS catalogues, $C$ is $g-r$ and $M$ is $r$.
The magnitude bin-size is the same  ($\delta r$ = 0.2\,mag) as in calculating $S_1$.
We adopt however a larger colour bin-size [$\delta (g-r)$ = 0.5\,mag] for calculating $S_2$ as this quantity is 
less sensitive to the colours of stars. 
In Fig.~\ref{cmdsp} we show the adopted colour-magnitude grid for some example 
spectrographs of three illustrative plates. 

The distribution of averaged values of $S_2$ in the $C$-$M$ space and in the LAMOST 
focal plane for all spectroscopic measurements with robust stellar 
parameters in LSS-GAC DR2 are shown in Fig.~\ref{f2d}. 
Bright and blue stars have higher averaged values of $S_2$ than faint and red stars.
The differences in the average values of  $S_2$ of the individual spectrographs are clearly visible, 
although the differences reduce significantly after averaging over all LSS-GAC plates. 
For example, Spectrographs~\#12 and \#13 near the edge of the
LAMOST focal plane have lower averaged values of $S_2$ (i.e. lower 
observing efficiencies) than Spectrographs~\#4 and \#8 near the centre of the focal plane.
The differences between the average values of  $S_2$ of the individual 
spectrographs are more pronounced for results
based on the XSTPS-GAC catalogue than those based on the APASS catalogue, 
with the fraction of faint (B/M) plates are larger  in the former case. 

\subsection{The final selection function} 

In the case that both $S_1$ and $S_2$ are evaluated using the same $C$ and $M$ bin sizes, 
then the final selection function, $S$, is simply given by the product 
of $S_1$ and $S_2$. However, as noted above, we use slightly different $C$ and $M$ bin sizes when 
evaluating $S_2$ compared to the bin sizes used for $S_1$. As such, we now define $C1$ and $M1$
for the colour-magnitude bin used to calculate $S1$. Similarly,  $C2$ and $M2$ is defined as 
the colour-magnitude bin used to calculate $S2$. The final selection function,
$S$, is given by
\begin{equation}
  S=\dfrac{N_{\rm PARAM}({\rm sp}, C2, M2)}{N_{\rm total}({\rm sp}, C2, M2)},
\end{equation}
where for each spectrograph ${\rm sp}$ and each $C2$ and $M2$ bin, 
$N_{\rm PARAM}({\rm sp}, C2, M2)$ is the number of stars having robust parameters and 
$N_{\rm total}({\rm sp}, C2, M2)$ is the total number of stars observed by LAMOST with target selection
function $S_1$                                                                                                   corrected. We thus have
\begin{equation}
N_{\rm total}({\rm sp}, C2, M2)={\sum\limits_{C1 \in C2}\sum\limits_{M1 \in M2} \frac{N_{\rm LAMOST}({\rm sp},C1,M1)}{S_1 ({\rm sp},C1,M1                                                                                                                                                                                                                                                                                                                                                                                             )}}.
\end{equation}

Example values of  selection functions $S_1$, $S_2$ and $S$ 
thus calculated for spectroscopic measurements in LSS-GAC DR2 are given in Table 1
for the purpose of illustration. The full results are available by contacting the authors (BQC, XWL) 
 and  will be included in the next release of LSS-GAC value-added catalogue, 
 i.e. LSS-GAC DR3 (Huang et al., in preparation). 

\section{Mock data test}

\begin{figure}
    \centering
\includegraphics[width=0.48\textwidth]{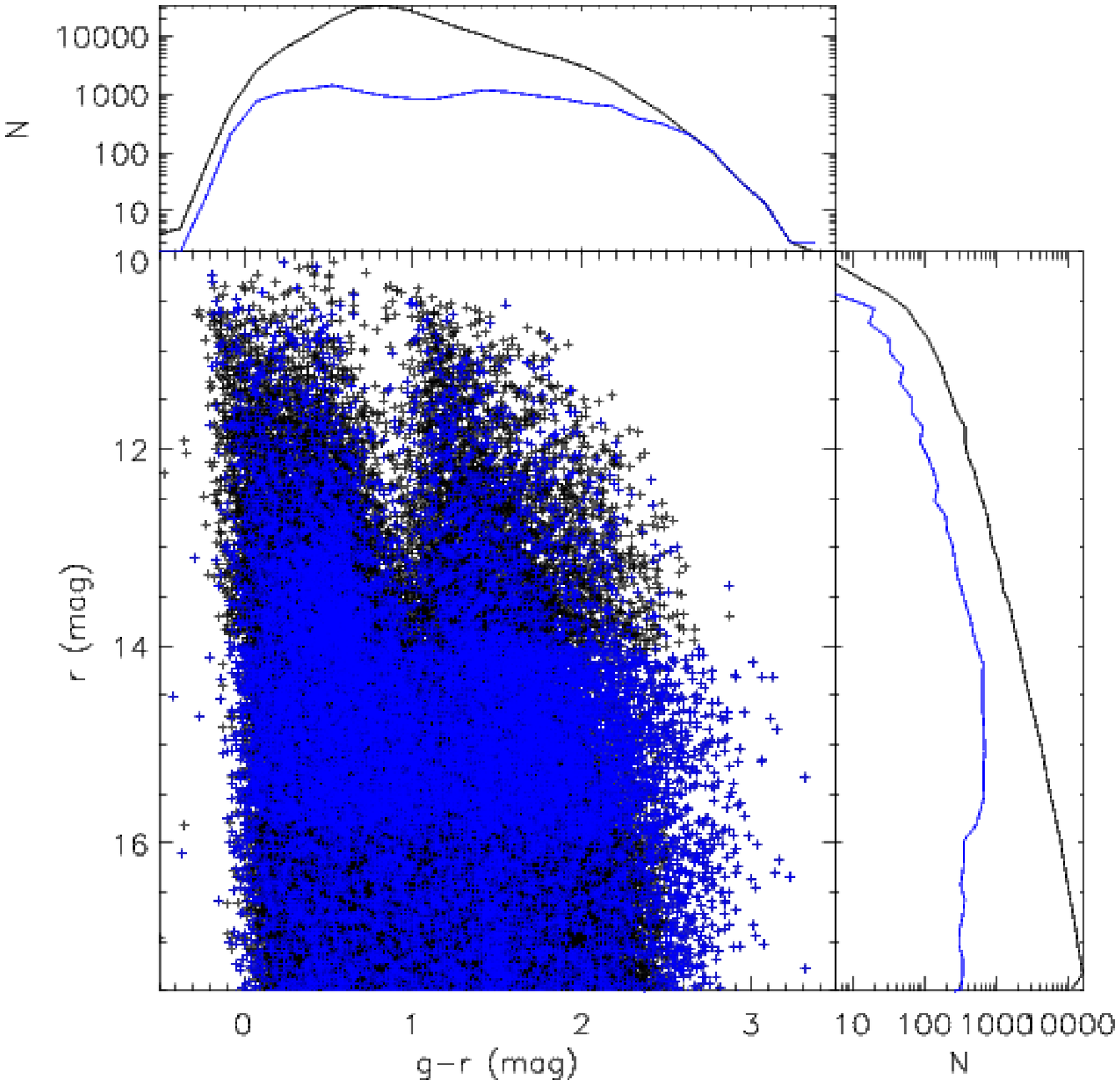}
\includegraphics[width=0.48\textwidth]{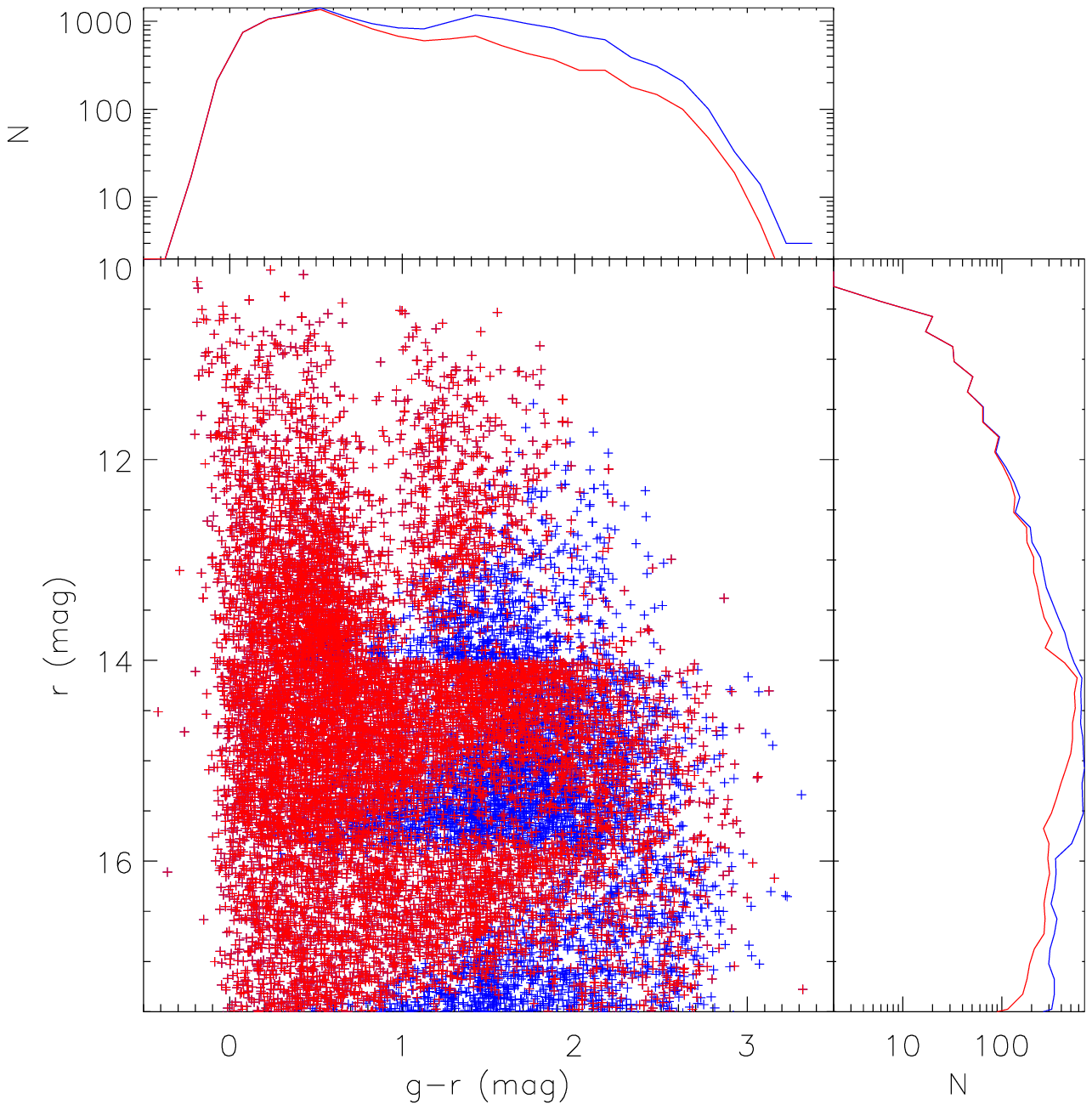}
  \caption{Top panel shows distributions in colour-magnitude plane of targets selected from the 
Besan\c{c}on simulated catalogue (black dots and histograms), 
using the same target selection algorithm as for LSS-GAC (blue dots and histograms). 
Selected targets with simulated `SNR' $>$ 10 (see text for details) are represented by 
red dots and histograms in the bottom panel.}
  \label{mkdata}
\end{figure}

\begin{figure}
    \centering
\includegraphics[width=0.48\textwidth]{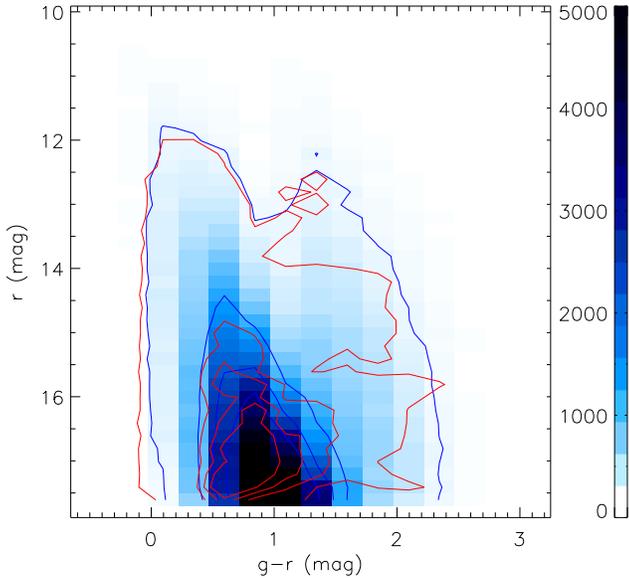}
  \caption{Hess diagram of the mock `photometric' sample  (blue scales and contours), compared 
to that given by the sample with `reliable' parameters after corrected for the selection biases (red contours).}
  \label{sfmk}
\end{figure}

\begin{figure*}
    \centering
\includegraphics[width=0.98\textwidth]{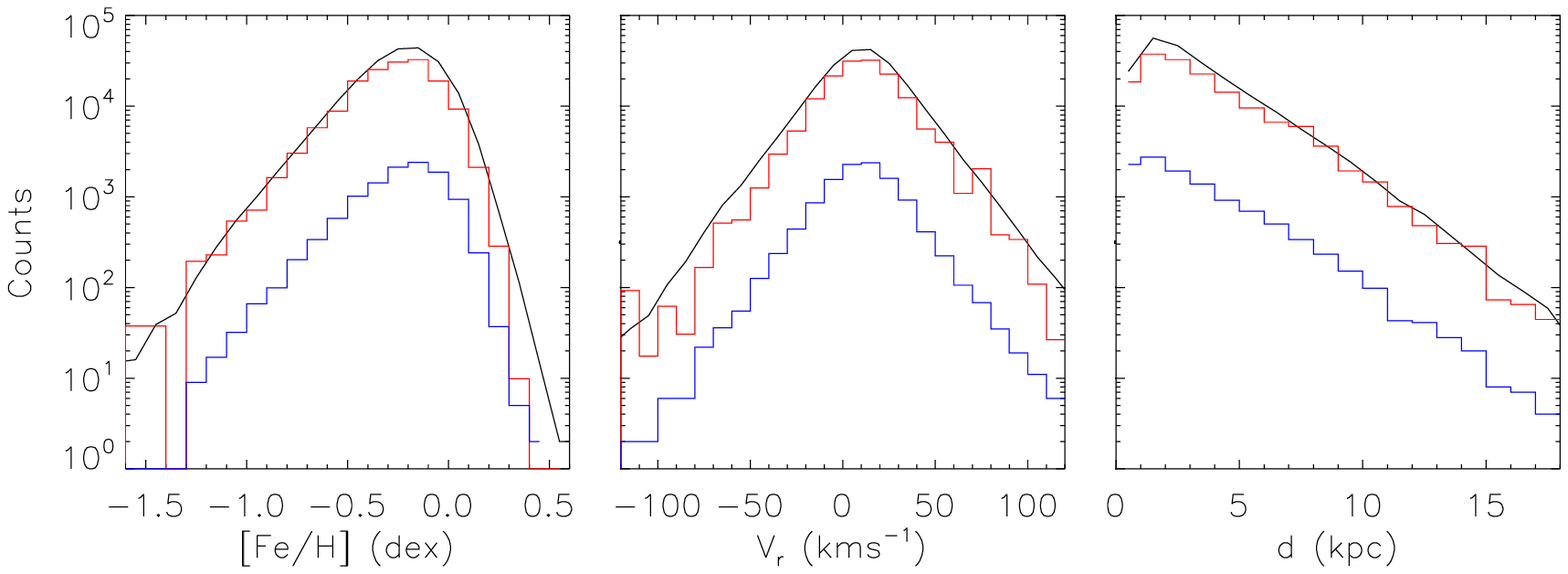}
  \caption{Distributions of metallicity [Fe/H], radial velocity $V_{\rm r}$ and 
 distance $d$ for the mock `photometric' sample (black histograms), the 
 sample with `reliable' parameters (blue histograms), and for the sample with
  `reliable' parameters after corrected for the selection effects (red histograms). 
 }
  \label{fvd}
\end{figure*}

In this Section we validate our method using a mock star catalogue.
For this purpose, we utilise the Besan\c{c}on
stellar population synthesis model \citep{Robin2003}
to generate a catalogue centred on the  GAC, ($l,~b$)~=~(180\degr, 0\degr). 
The three-dimensional extinction maps from \citet{Chen2014} are used to add
extinction to stars in the catalogue.  Taking  the simulated catalogue 
as an observed one, we 
select targets using the same algorithm as 
adopted for the LSS-GAC target selection. We artificially define 
a field centred on  ($l,~b$)~=~(180\degr, 0\degr).  
Within this field, four plates, one VB,  two B  and one M plates, 
are generated, containing 4 000, 3 888, 3 914
and 3 950 stars, respectively. In Fig.~\ref{mkdata} we plot the
colours and magnitudes of the selected targets, along with those of 
all stars in the full `photometric' catalogue. 

To define a sample of stars with `reliable stellar parameters',
we artificially add SNRs to all of the selected targets.
To simulate the quality of stellar spectra in real LAMOST observations, 
we randomly select three plates from LSS-GAC: 
a VB plate HD213239N421743V01, a B plate GAC101N22B1 and an M plate GAC091N33M1. 
For each spectrograph of each plate, the distribution of \bsnr\ of stars in the selected plates is fitted as 
a function of colour $g-r$ and magnitude $r$, as,
\begin{equation}
  {\rm SNR}=a_1+a_2(g-r)+a_3(g-r)^2r+a_4(g-r)r+a_5r+a_6r^2,
\end{equation}
where $a_1$, $a_2$, ..., and $a_6$ are the coefficients. The fitting
is then applied to all spectrographs of simulated plates 
to assign `real' SNRs for those selected and get `observed' stars. 
We ignore here the effects of the dead, saturated or contaminated 
fibres and the uncertainties induced by the LSP3 pipeline. 
Assuming that the LSP3 pipeline is able to deliver robust 
stellar parameters for all stars with a `SNR' $>$ 10,
the sample of stars with robust parameters is simply defined by 
stars with a `SNR' $>$ 10.
In Fig.~\ref{mkdata}, stars in this sample with `robust' parameters  are also overplotted. 

We now have a `photometric' sample generated with the Besan\c{c}on model, 
a sample of stars selected with the same target selection algorithm as for the LSS-GAC and get `observed', 
and a sample with  `reliable'  parameters by adding artificially `SNRs' to their `observed' spectra. 
The values of selection function for each star with `reliable' parameters are then calculated following the 
procedure introduced in the former Section.

The sample with `reliable' parameters is then corrected for 
selection biases using the calculated selection function. In Fig.~\ref{sfmk}, 
we compare the resultant colour and magnitude distribution
of the sample with  `reliable' parameters after correcting for the selection effects with the distribution of  
the underlying population (i.e. the `photometric' sample). 
Overall the agreement is quite good. 
For regions of extreme red colours and faint magnitudes,
given the small number of stars with `reliable' parameters, we 
 are not able to perfectly recover the CMD of the underlying stellar population.
 
We have also tested our selection function results for 
 the metallicity,  radial velocity and star count distributions. 
Comparisons of those distributions as given the sample with 
`reliable' parameters after corrected for the selection effects and those of the
underlying population are shown in Fig.~\ref{fvd}. 
Overall, the agreement is good.
We note that for stars of extreme parameters, such as those of very high metallicities, 
very high radial velocities, or very far distances, due to the large statistical uncertainties, 
the two sets of distributions do not match well, but are still consistent with each other.
Thus the selection function presented in the current work is a powerful tool for the 
 studies of Galactic chemistry and dynamics.
 
\section{Applications of the selection function}

\begin{figure}
    \centering
\includegraphics[width=0.45\textwidth]{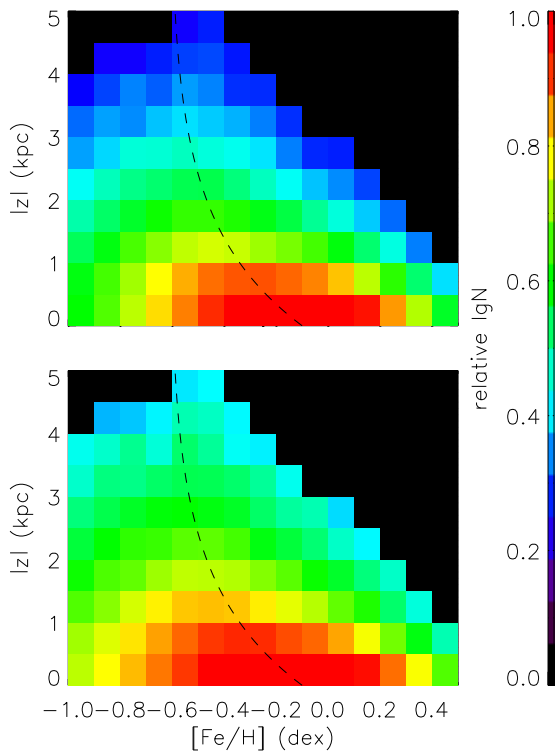}
  \caption{Distribution of stellar number density in the metallicity [Fe/H] and height from 
the Galactic mid-plane $|z|$ from a sample of LAMOST F-type stars.
The density is shown on a logarithmic scale, with peak value 
normalized to unity. The dashed lines show Eq.~(11) of \citet{Chen2017}.
The top and bottom panels show results before and after applying the selection function corrections, respectively.}
  \label{mdf}
\end{figure}

\begin{figure*}
    \centering
\includegraphics[width=0.88\textwidth]{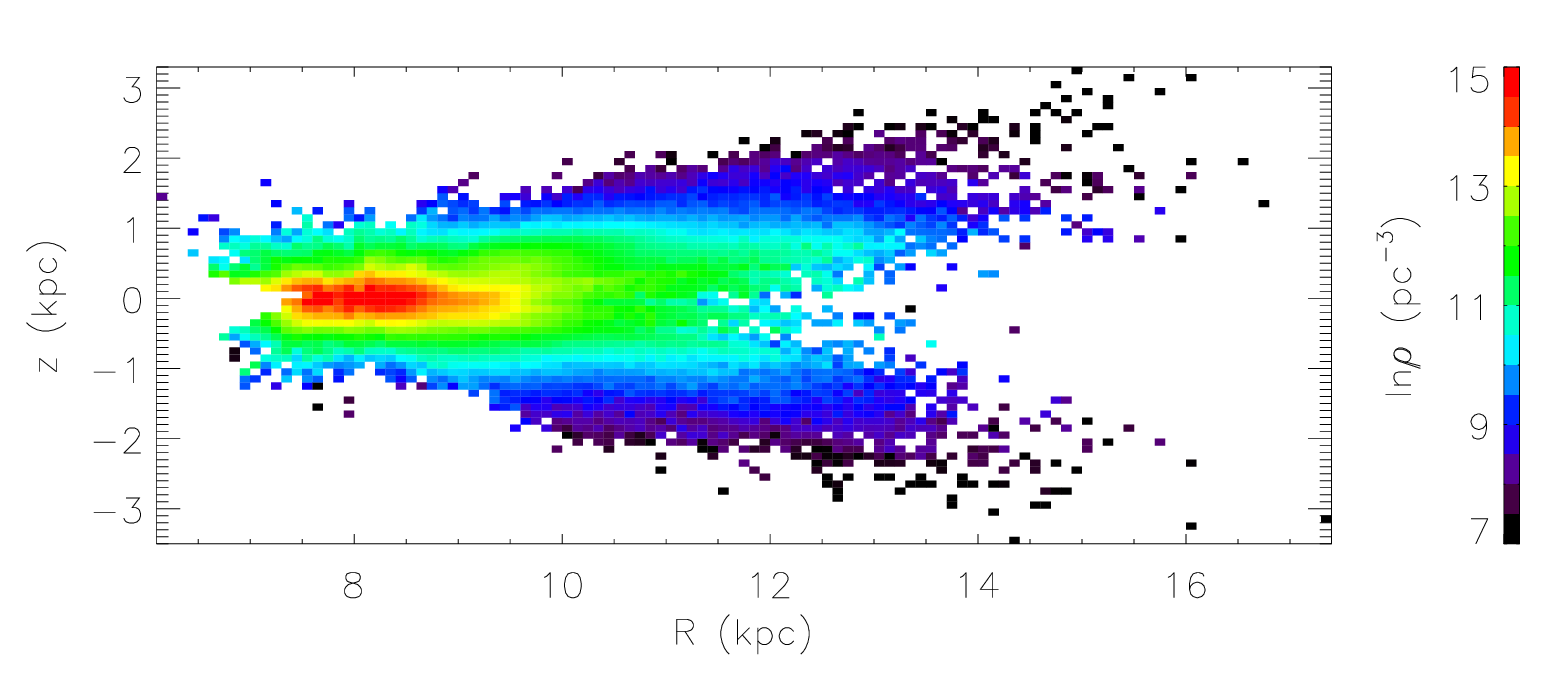}
  \caption{Stellar density distribution in the $R$ - $z$ plane 
  deduced from a sample of  LAMOST F-type stars after corrected for  selection biases. 
  The colour encodes the mean ln\,$\rho$ value in each $R$-$z$ bin. }
  \label{ndis}
\end{figure*}

In this Section we give two simple applications of our selection function. 
We select a sample of F-type stars from the internal release of  LSS-GAC DR2
with effective temperature and surface gravity cuts,
6000 $<$ $T_{\rm eff}$ $<$ 6800\,K
and 3.8 $<$ log\,$g$ $<$ 5.0\,dex. 
The internal release of LSS-GAC DR2 includes all observations from the initiation of the survey 
up to 2016 June. The selected F-type star sample contains 713 016 spectroscopic measurements.

\subsection{The metallicity distribution}

In Fig.~\ref{mdf}, we plot the distribution of stellar number density in the plane of 
metallicity [Fe/H] and height from the Galactic mid-plane $|z|$  of this sample
for the metallicity and height ranges, $-$1 $< {\rm [Fe/H]} <$ 0.5\,dex and 0 $< |z| < $ 5\,kpc.
The peak value of the distribution is normalized to unity. 
The upper and bottom panels show 
respectively the results derived before and after applying 
the selection function corrections deduced in the current work.
When plotting the distributions, we have discarded bins containing fewer than
10 stars. For different $|z|$ slices, the variations of peak metallicities as a function of $|z|$ is consistent
with the fit given by Eq.~(11) of \citet{Chen2017},
a fit obtained using main sequence turn off stars selected from
LSS-GAC DR2 \citep{Xiang2015}. 
The variations of peak metallicities as $|z|$  
are quite similar before and after applying the 
selection function corrections, 
suggesting that the LSS-GAC selection function
has only a marginal effect on the metallicity peak distributions. 
This is in consistent with the recent work by \citet{Nandakumar2017}.

Amongst the individual $|z|$ slices, both the metallicity dispersion and skewness vary. 
As $|z|$ increases, the metallicity dispersions decrease while the skewness increase,
which imply the star formation and radial migrations history 
of the Galactic thin and thick disks \citep{Sellwood2002, Schonrich2009, Hayden2015, Loenman2016}. 
A detailed analysis of the Galactic disk metallicity distribution based on the LAMOST
main sequence turn off stars will be presented 
in a separate work (Wang et al., in preparation).
Compared to the distributions after the selection effect corrections, 
those before the selection function corrections
have smaller dispersions and larger skewness.
This is likely to be caused by the fact that stars further away (of larger $|z|$) are 
fainter and thus suffer from
larger selection effects than those nearby ones.  

\subsection{The stellar number density distribution}

Fig.~\ref{mdf} also shows for different metallicity bins how the stellar number density 
varies with $|z|$. It is clear that the number densities of metal-poor populations 
decrease more slowly than those of the metal-rich ones, in other words, the metal-poor populations
have larger scale heights.
There is no doubt that applying the selection function corrections is very important 
for this type of study. Without the corrections, the scale heights derived 
will be systematically underestimated.
This is again largely due to the fact that stars further away suffer from larger selection effects than those nearby. 

With the selection function corrections presented here, one can thus examine the underlying 
stellar number density distributions using the LAMOST spectroscopic samples. 
We give an example here using the F-type star sample. 
From stars observed in each spectrograph of each plate,
one can simply derive the stellar number density using,
\begin{equation}
  \rho_j = \dfrac{\Sigma_{i, j}(\frac{1}{S_{i,j}})}{V_{j}},
\end{equation}
where $i$ is the index of stars in a distance bin of index $j$, 
and $V_{j}$ is the volume of the $j$th distance bin.
We adopt the centre $l$ and $b$ for each spectrograph  and convert ($l,~b,~d$) into 
the Galactocentric cylindrical coordinates ($R,~z,~\phi$) similar as in \citet{Bovy2012} and
\citet{Liu2017}.
The resultant number density is then averaged in the $R$ and $z$ plane. 
The results are presented in Fig.~\ref{ndis}. 
The Figure displays a remarkable shape of the Galactic disk, 
very similar to that seen in an edge-on external
disk galaxy, for Galactic radius $R$ between 5 and 16\,kpc. Note that here we have assumed that the 
photometric sample from which LAMOST targets are drawn is complete. 
In addition, we have also ignored the possible variations of 
the absolute magnitudes of F-type stars.  Any such variations, 
coupled with the varying interstellar extinction, affect the lower and upper completeness 
distance limits for the individual lines of sight lines, effects that one 
must take into account when studying the Galactic structure \citep{Chen2017}.   
More quantitative analysis will be presented in a separate work ( Chen et al., in preparation).

\section{Summary}

In this paper, we have discussed in detail the selection function of LSS-GAC spectroscopic survey and
presented corrections for all spectroscopic measurements with reliable parameters in the LSS-GAC DR2. 
The selection function determines how representative the final spectroscopic catalogue is 
compared to the underlying stellar population of the Milky Way. It is a powerful tool for the studies of the 
Galactic chemistry and structure problems.

We divide the selection function into two parts.
The first part, quantified by $S_1$, characterise the LSS-GAC target selection strategy.
The LSS-GAC target selection is based on stellar magnitudes 
and colours, using photometric data from the XSTPS-GAC, supplemented by PPMXL and 2MASS photometry
for the VB survey. Based on the photometric data of XSTPS-GAC and APASS DR9,
we calculate $S_1$  in the $C$ and $M$ space
for each spectrograph of each LSS-GAC plate. 
The second part, quantified by $S_2$, characterise the selection effects due to the observational quality, 
data reduction and parameter determination.
We select from LSS-GAC DR2 a commonly used sample that contains stars with 
reliable stellar parameters. Values of $S_2$ of the sample are calculated 
for each $C$-$M$ bin and for each spectrograph of each plate.
The full selection function $S$ can then be calculated from 
$S_1$ and $S_2$. Example values of selection function corrections are listed in Table~1. 
The full results are available upon request by email, and  
will be included in the next release of LSS-GAC value-added catalogue.

We test our method using mock data. The test shows that the selection function corrections presented here can 
successfully recover the distributions of colours, magnitudes, 
metallicities, radical velocities, as well as number counts of the underlying stellar populations.
Finally we present two simple applications of our deduced selection function corrections. 
The selection function presented in the current work provide a better insight of the properties of LSS-GAC 
and the resulted value-added catalogue, and can be used to study a variety of problems of the Milky Way galaxy that rely on proper corrections for the selection biases in the LSS-GAC spectroscopic dataset.

\section*{Acknowledgements}

We thank our anonymous referee for helpful comments that improved
the quality of this paper.
This work is partially supported by National Key Basic Research Program of China
2014CB845700,  China Postdoctoral Science Foundation 2016M590014 and National
Natural Science Foundation of China U1531244. 
The LAMOST FELLOWSHIP is supported by Special Funding for Advanced Users,
budgeted and administrated by Center for Astronomical Mega-Science, Chinese
Academy of Sciences (CAMS). 

This work has made use of data products from the Guoshoujing Telescope (the
Large Sky Area Multi-Object Fibre Spectroscopic Telescope, LAMOST). LAMOST
is a National Major Scientific Project built by the Chinese Academy of
Sciences. Funding for the project has been provided by the National
Development and Reform Commission. LAMOST is operated and managed by the
National Astronomical Observatories, Chinese Academy of Sciences.

This research was made possible through the use of the AAVSO Photometric 
All-Sky Survey (APASS), funded by the Robert Martin Ayers Sciences Fund.


\bibliographystyle{mn2e}
\bibliography{selfun}

\label{lastpage}
\end{document}